\documentclass[12pt,oneside]{article}
\usepackage[centertags]{amsmath}
\usepackage{amssymb}
\usepackage{amsthm}
\usepackage{bm,url}
\usepackage{mathptmx}       
\usepackage{helvet}         
\usepackage{courier}        
\usepackage{type1cm}        
\usepackage{amsfonts}
\usepackage{makeidx}         
\usepackage{graphicx}        
\usepackage{multicol}        
\usepackage[bottom]{footmisc}
\usepackage{multirow}
\usepackage{graphics}  

\makeindex             
\newcommand{\cub}{\textsc{cub\,}}

\begin{document}
\title{CUB models: a preliminary fuzzy approach to heterogeneity}
\author{E. Di Nardo\footnote{Department of Mathematics \lq\lq G. Peano\rq\rq, Via Carlo Alberto 10,
University of Turin, 10123, I-Turin, elvira.dinardo@unito.it},  R. Simone\footnote{Department of Political 
Sciences, University of Naples Federico II, Via Leopoldo Rodin\`o, 22, 80133 I-Naples, rosaria.simone@unina.it}}
\date{ }
\maketitle
\begin{abstract}
In line with the increasing attention paid to deal with uncertainty in ordinal data models, we propose to combine Fuzzy models with \cub models within questionnaire analysis. In particular, the focus will be on \cub models' uncertainty parameter and its interpretation as a preliminary measure of heterogeneity, by introducing membership, non-membership and uncertainty functions in the more general framework of Intuitionistic Fuzzy Sets. Our proposal is discussed on the basis of the Evaluation of Orientation Services survey collected at University of Naples Federico II.
\end{abstract}
\smallskip
{\bf keywords:} \cub models, Fuzzy Set, Membership Functions, Uncertainty.
\smallskip
\section{Introduction} \label{intro}
In ordinal data models, the understanding of the mechanism that leads respondents to produce an evaluation out of a latent perception comes with the need of distinguishing between randomness and uncertainty, in all its different sources. \\
Following the \cub models rationale \cite{DelPic2005,Ianp12}, a respondent marks a score on an ordinal scale according to a data generating process which is basically structured as the combination of two components: the feeling, responsible for the level of agreement/pleasantness towards the item under investigation, and the uncertainty, accounting for the overall nuisance affecting a fully meditated response (laziness, inherent difficulties in understanding the question, ignorance of the subject, etc), that is fuzziness. \cub models are then defined as a two-component mixture distribution: a shifted Binomial for feeling and a discrete Uniform for uncertainty. Such a probabilistic way to describe the inherent indeterminacy of human decisions helps make \cub models a valid alternative in the scenario of categorical data models. However, an effective questionnaire analysis requires the simultaneous examination of all the items, and \cub models are currently lacking of a multidimensional setting so that only an item-by-item investigation can be run (first steps in this direction have been achieved in \cite{Ferrari, Corduas}). Fuzzy Sets Theory instead accounts for uncertainty in questionnaire analysis by evaluating respondents on more items \cite{Lalla, Zadeh}. 

As underlined in \cite{Marasini}, the Fuzzy approach involves a certain level of subjectivity in the choice of membership functions' shapes, which is both a drawback and an advantage. Conversely, \cub distributions provide a more objective quantification of the uncertainty component, though are less flexible since the probability structure is given. More accurate specifications of the uncertainty are available in both frameworks. For instance Intuitionistic and Hesitant Fuzzy Sets (IFS and HFS, respectively) are considered in \cite{Marasini} aiming to  separate the non-membership modeling from the hesitation functions. The same happens in \cub models, where alternative choices for the uncertainty distribution have been proposed \cite{VarCUB}. Aware that Fuzzy and \cub models are structurally different paradigms oriented to model uncertainty, our goal is to run a first attempt of merging the potentiality of \cub models within Fuzzy Sets Theory by proposing a variation of a well-stated choice of the membership function recently discussed in \cite{Zani}. Our approach results in suitably weighting the various membership degrees whenever the distribution presents a considerable level of heterogeneity. On the other hand, the proposed methodology endeavors to lead the way to a multidimensional analysis with \cub models.
The latent phenomenon we shall consider is the satisfaction of respondents: in this regard, we shall also model non-membership and uncertainty functions as prescribed in the more general framework of IFS \cite{Marasini}. Our proposal is discussed on the basis of the Evaluation of Orientation Services surveys collected at University of Naples Federico II over different waves (see Section \ref{Ver}). The whole analysis has been run within the \texttt{R} environment.

\section{CUB models}
\cub is an acronym that stands for \textit{Combination of a Uniform and a shifted Binomial} random variables, since the \cub distribution consists in the following two-component mixture of parameters $\pi, \,\xi$:
\begin{equation}
\label{prcub}
Pr\big(R=r \mid \pi, \xi \big) =\pi\,b_r(\xi)+(1-\pi)\,h_r\,, \quad r=1,2,\dots,m\,,
\end{equation}
where $b_r(\xi)$, $r=1,2,\dots,m$ for a given $m>3$ denotes the shifted Binomial distribution of parameter $\xi$:
\begin{equation}
\label{shifbin}
b_r(\xi) = \binom{m-1}{r-1}\xi^{m-r}(1-\xi)^{r-1}, \quad r=1,2,\dots,m\,,
\end{equation}
and $h_r = \frac{1}{m}$ is the discrete Uniform distribution on the given support. The parameter $\xi$ is referred to as the feeling parameter since $1-\xi$ measures the preference of a category over the preceding ones in a pairwise comparison \cite{Del00a}. The parameter $\pi$, instead, is the uncertainty parameter since $1-\pi$ is the mixing proportion of the Uniform distribution. This choice represents the least informative situation and hence $1-\pi$ aims at catching the level of heterogeneity in the distribution, thus measuring respondents' attitude towards a non-meditated/Fuzzy behavior.
For our discussion, it is worth of interest that the uncertainty parameter $\pi$ can be preliminarily estimated as an heterogeneity measure starting from the relation \cite{Iannario}:
\begin{equation}
\label{bestpaicub}
\mathcal{G}_{\cub} = 1 - \pi^{2}(1-\mathcal{G}_{SB})
\end{equation}
where 
\begin{equation}
\label{Gini}
\mathcal{G} = \frac{m}{m-1}\bigg( 1 - \sum_{r=1}^{m}f_r^2\,\bigg)
\end{equation}
denotes the normalized Gini heterogeneity index for a given frequency distribution $(f_1,\dots,f_m)$. However, the resulting estimate might result quite biased \cite{Iannario}, especially for distributions with extreme feeling. This is why we shall consider the maximum likelihood estimators (ML) of parameters\footnote{We underline that the uncertainty parameter will be considered as an a heterogeneity measure even if we shall rely on ML estimates rather than on \eqref{bestpaicub}.} obtained by running the Expectation-Maximization (EM) algorithm \cite{DelPic2005} as implemented in the \texttt{R} package \cub \cite{CUBpackage}. 
\section{Fuzzy system models}
Fuzzy Sets Theory originates with the work of Zadeh \cite{Zadeh} and since then it has been the focus of several research purposes. One of its main promising application fields is within social measurements achievable with questionnaire analysis \cite{Lalla, Marasini}, which partially motivates the following analysis. 

For a given universe of discourse $X$, a Fuzzy set $A$ consists of a subset of $X$ endowed with a membership function $\mu_A$ 
measuring the degree of membership to the set $A,$ that is, 
\begin{equation*}
\mu_A :\, X \longrightarrow [0,1], \qquad x \longmapsto \mu_A(x),
\end{equation*}
in such a way that $\mu_A(x) =1$ if and only if $x$ is certainly an element of $A$, while $\mu_A(x) =0$ if and only if $x$ is certainly not. Additionally, the rationale of IFS \cite{Marasini} is to supply the analysis with a non-membership function:
\begin{equation*}
\nu_A :\, X \longrightarrow [0,1], \qquad  x \longmapsto \nu_A(x),
\end{equation*}
expressing the complementary assessment of the level of non-membership of an element $x$ to the Fuzzy set $A$, in such a way that if $\nu_A(x) =1$, then $x$ is certainly not an element of $A,$ and more generally:
$$  0 \leq \mu_A(x) + \nu_A(x) \leq 1.$$
Then, a measure of the residual indecision about the statement $x\in A$ is given by the Fuzzy uncertainty function:
\begin{equation}
\label{UncFunction}
u_A(x) = 1 - \mu_A(x) - \nu_A(x).
\end{equation}
Questionnaire analysis generally involves a simultaneous examination of all the items in order to yield an overall evaluation of the latent phenomenon under investigation: for our purposes, we consider satisfaction for $n$ respondents. The standard defuzzification procedure consists in computing \textit{membership} and \textit{non-membership scores} by using crisp synthetic indicators of the overall degree of membership/non-membership to the set $A$ of the ensemble of satisfied respondents. 

Given an item-by-item analysis, the following aggregation strategy has been considered. Assume that a questionnaire is designed with $K$ items, say $X_1,\dots,X_K$. Within IFS, a standard approach is to consider the IWAM (Intuistionistic Weighted Aggregator Mean) \cite{Marasini} both for membership and non-membership functions: 
\begin{equation}
\label{IWAM}
<\mu_A(\mathbf{r}_j), \nu_A(\mathbf{r}_j) >  \,= \, <  \sum_{k=1}^K w_k \, \mu_A(r_{j,k}), \,\sum_{k=1}^K w_k \, \nu_A(r_{j,k}) >,
\end{equation}
where $\{w_1, \ldots, w_K\}$ is a given system of weights and $\mathbf{r}_j =(r_{j,1},r_{j,2},\dots, r_{j,K})$ is the vector of observations
given by the $j$-th respondent, consisting in rates $r_{j,k}$ to the $k$-th item.  

Following \cite{Marasini}, we shall consider each aggregated value $<\mu_A(\mathbf{r}_j), \nu_A(\mathbf{r}_j) >$ as an IFS Fuzzy singleton $ <j, \mu_A(\mathbf{r}_j), \nu_A(\mathbf{r}_j) >$ and then compute again the IWAM aggregator (\ref{IWAM}) with equal weights $w_k = \frac{1}{n}$, yielding to the final scores:
\begin{equation}
\label{score}
<\bar{\mu},\bar{\nu}> = <\frac{1}{n}\sum_{j=1}^n \mu_A(\mathbf{r}_j),  \frac{1}{n}\sum_{j=1}^n \nu_A(\mathbf{r}_j) >.
\end{equation}
Accordingly to \eqref{UncFunction}, the \textit{uncertainty score} is the overall residual degree of indeterminacy:
\begin{equation}
\label{uncertaintyScore}
\bar{u} = 1 - \bar{\mu} -\bar{\nu}.
\end{equation}
In \cite{Zani}, the weights are computed by using the logged inverse of the Fuzzy proportions of the achievement of the target (the respondents' satisfaction) for each item:
\begin{equation}
\label{FuzzyProp}
g(X_k) = \frac{1}{n}\sum_{j=1}^n \mu_A(r_{j,k}), \,\, \hbox{for}\,\,\,\, k=1, \ldots, K
\end{equation}
and then normalizing, as 
\begin{equation}
\label{pesiZani}
w_k = \ln\bigg( \frac{1}{g(X_k)}\bigg) \bigg/ \sum_{l=1}^{K} \ln\bigg( \frac{1}{g(X_l)}\bigg), \,\, \hbox{for}\,\,\,\, k=1, \ldots, K. 
\end{equation}
However, since the weights should be larger for the more explanatory items \cite{Marasini}, in \eqref{FuzzyProp} we consider the fuzzy proportions of uncertainty functions \eqref{UncFunction} rather than of the membership functions:
\begin{equation}
\label{FuzzyPropUF}
g(X_k) = \frac{1}{n}\sum_{j=1}^n u_A(r_{j,k}),\,\, \hbox{for}\,\,\,\, k=1, \ldots, K.
\end{equation} 
With this choice, items with a larger uncertainty (in the sense of \cub models) should result less informative and crucial in determining the overall satisfaction of respondents.  

\section{Membership functions and uncertainty}
\label{MF}
In the following, we shall consider an ordinal scale with an odd number of categories and with an indifference point $i_p$ located at the mid category. The ordinal scale is oriented as such ``the greater the score, the higher the feeling'', that is, we consider a positive relation between the variable and the scale. We shall also assume that the scale has equidistant categories, say $1,2,\dots,m$, so that a rate $r=1$ corresponds to the most negative choice; conversely, $r=m$ corresponds to the extreme positive answer.
In order to propose a Fuzzy composite indicator for customer satisfaction, Zani \textit{et al.} consider the well-known membership function
\cite{Cerioli}:
\begin{equation}
\label{Zani}
\mu_A(r) =
\begin{cases}
0, & \quad 1 \leq r \leq l_b, \\
\mu_{A}(r-1) + \dfrac{F(r) - F(r-1)}{1-F(l_b)}, &  \quad  l_b < r < u_b, \\
1, &\quad u_b \leq r \leq u_b 
\end{cases}
\end{equation}
where $F(r)$ denotes the empirical distribution function, $l_b$ ($u_b$, resp.) is a fixed lower (upper, resp.) bound to threshold the categories corresponding to negative (positive) scores. Customarily, $l_b$ is the least negative choice while $u_b$ is chosen to be the second to last positive choice. From (\ref{Zani}), the membership degree $\mu_A(r)$ is updated with respect to $\mu_{A}(r-1)$ with the relative frequency of category $r$ normalized to the relative frequency of answers that are not considered negative choices. However, the greater the heterogeneity is in the whole distribution, the less meaningful the relative frequency should be considered as membership degree. 

We propose to modify (\ref{Zani}) as follows: 
\begin{equation}
\label{CUBZani}
\mu_A(r) =
\begin{cases}
0, & \quad 1\leq r \leq l_b =i_p-1,\\
\dfrac{1-\hat{\pi}}{m}, & \quad r=i_p, \\
\mu_{A}(r-1) + \hat{\pi}\,\dfrac{F(r) - F(r-1)}{F(u_b-1) - F(i_p)}, &  \quad  l_b < r < u_b, \\
1, &\quad u_b \leq r \leq m,
\end{cases}
\end{equation}
in such a way that:
\begin{description}
\item[{\it i)}] the updating of category $r$ is penalized with the overall estimated uncertainty $\hat{\pi}$, as results from a \cub model fitted to the data;
\item[{\it ii)}] the indifference point $i_p$ is highlighted with the heterogeneity level, as measured by a \cub model fitted to the data;
\item[{\it iii)}]  the frequency of category $r$ is normalized taking into account the set of positive non-crisp choices.  
\end{description} 
According to the IFS approach, we define the non-membership function by similar arguments as:
\begin{equation}
\label{nmf}
\nu_A(r) =
\begin{cases}
0 & \quad i_p < \,r \leq m,  \\
\dfrac{1 - \hat{\pi}}{m}, & \quad r = i_p,\\
\nu_{A}(r+1) + \hat{\pi}\,\dfrac{F(r) - F(r-1)}{F(l_b)-F(1)},&  \quad  1 < r \leq l_b=i_p-1, \\
1 &\quad r = 1. 
\end{cases}
\end{equation}
Finally, the uncertainty function will be given as the residual fuzziness:
\begin{equation}
\label{uncertainty}
u_A(r)  = 1 - \mu_A(r) - \nu_A(r), \quad r=1,\dots,m.
\end{equation}
Let us motivate more deeply our proposal. First, let us consider the insertion of the indifferent point $i_p.$
The choice of giving an \textit{ad-hoc} assignment to the membership and non-membership degrees of $i_p$ is motivated by the following argument: in the theoretical scenario of no uncertainty (that is, as $\hat{\pi} \rightarrow 1$), ratings corresponding to $i_p$ should receive a null degree both of membership and non-membership, because in this case one should be able to perfectly classify respondents (those giving a rate higher than $i_p$ belong to the set of satisfied respondents, with increasing degrees, those marking a score lower than $i_p$ do not). Hence, the indifference expressed by rating $i_p$ should be intended as {\it all choices are considered equivalent} for the respondent. This justifies the equality of both membership and non-membership degrees to the value $\frac{1-\hat{\pi}}{m},$ corresponding to the part of the \cub mixture expressing the Fuzzy behavior. Secondly, as the overall uncertainty decreases (that is, the more $\hat{\pi}$ approaches $1$), the more the non-membership function increases towards $1$ by moving from the indifference point to the first category.
For increasing heterogeneity (that is, as $\hat{\pi} \rightarrow 0$), from \eqref{CUBZani} and \eqref{nmf} we have:
$$\mu_A(i_p) = \nu_A(i_p) = \dfrac{1}{m},$$
which in turn implies:
$$\mu_A(r) = \dfrac{1}{m},\; r= i_p, \dots, u_b -1, \qquad \nu_A(s) = \dfrac{1}{m}, \; s= 2, \dots, i_p. $$
That is, the \textit{intermediate} categories are equally assigned a degree of membership, yielding a sort of trimmed uniformity among categories. Finally, referring to the normalization constant of the updating frequency, let us focus our attention on $F(u_b-1) - F(i_p)$ in \eqref{CUBZani}. As $\mu_A(r)=1$ for $r \geq u_b$, these categories are certainly associated with satisfaction. Hence, the shades of membership across intermediate positive categories should be computed starting from the indifference point and excluding the categories being assigned crisp membership degrees. Symmetric arguments lead to the choice of the normalization constant $F(l_b) - F(1)$ in \eqref{nmf}.
\section{A real case study}
\label{Ver}
The survey on Evaluation of Orientation Services has been collected at University of Naples Federico II from 2002 to 2008, across all the 13 Faculties, aiming at measuring the global satisfaction toward the service \footnote{Data are available at  \text{\url{http://www.labstat.it/home/research/resources/cub-data-sets-2/}.}}. On a balanced 7 point Likert scale: $1$=extremely unsatisfied, $2$=very unsatisfied, $3$=unsatisfied, $4$=indifferent, $5$=satisfied, $6$=very satisfied, $7$=extremely satisfied, the following items were questioned:
\begin{itemize}
\item satisfaction on the acquired information (\texttt{informat});
\item evaluation of the willingness of the staff (\texttt{willingn});
\item adequacy of time-table of opening-hours (\texttt{officeho});
\item evaluation of the competence of the staff (\texttt{compete});
\item global satisfaction (\texttt{global}).
\end{itemize}
The present discussion will concern the data collected in 2002, consisting of $2179$ observations. The ML estimates for parameters of a \cub model fitting the data are summarized in Table \ref{tab:param}: \texttt{officeho} is the item with the highest estimated uncertainty, followed by \texttt{informat} and then by \texttt{compete}. Overall, there is a moderately low level of uncertainty and an extreme positive feeling.
\begin{table}[h!]
\caption{\cub parameter estimates}
\label{tab:param}       
\centering
\begin{tabular}[pos=c]{cccccc}
\hline\noalign{\smallskip}
&  \texttt{informat} & \texttt{willing} & \texttt{officeho} & \texttt{compete} &  \texttt{global} \\
\noalign{\smallskip}\hline\noalign{\smallskip}
$\hat{\pi}$\;  & 0.7936 & 0.8567 & 0.6802 & 0.8022 &  0.8684 \\
$\hat{\xi}$ \; &   0.1809 & 0.1167 & 0.1971 & 0.1638 & 0.1714 \\
\noalign{\smallskip}\hline\noalign{\smallskip}
\end{tabular}
\end{table}

Next, we compare the membership function \eqref{Zani} exploited in \cite{Zani} with the \cub models adaptation \eqref{CUBZani} for all the investigated items.

\begin{table}[h!]
\caption{Membership Functions \eqref{Zani} and \eqref{CUBZani}}
\label{tab:mf}       
\centering
\begin{tabular}[pos=c]{ccccccc}
\hline\noalign{\smallskip}
Item & $\mu_A(r)$ as in &  $R \leq 3 $ & $ R = 4$ & $R = 5$ & $R = 6$ & $R =7 $  \\
\noalign{\smallskip}\hline\noalign{\smallskip}
\multirow{2}{*}{\texttt{informat}} & \eqref{Zani} & 0  & 0.0796 & 0.3483 & 0.6680   & 1 \\
 & \eqref{CUBZani} & 0 &  0.0295&  0.3919  & 0.8230  & 1  \\
 \hline \multirow{2}{*}{\texttt{willingn}} & \eqref{Zani} & 0 & 0.0453 & 0.2111 & 0.5154& 1\\
 & \eqref{CUBZani}& 0 &0.0205 & 0.3226 & 0.8772   & 1 \\
 \hline \multirow{2}{*}{\texttt{officeho}} & \eqref{Zani} & 0 & 0.1205 & 0.4081 & 0.6776 & 1\\
 & \eqref{CUBZani} & 0 &  0.0457 & 0.3969 & 0.7259&   1 \\
  \hline \multirow{2}{*}{\texttt{compete}} & \eqref{Zani} & 0 & 0.0811 & 0.3077 & 0.6380 & 1 \\
  & \eqref{CUBZani}& 0 & 0.0283&  0.3547 & 0.8305     &   1 \\
  \hline \multirow{2}{*}{\texttt{global}} & \eqref{Zani}  & 0 & 0.0726 & 0.2978 & 0.6673& 1\\
  & \eqref{CUBZani} & 0 &  0.0188 & 0.3477& 0.8872 &   1 \\
\noalign{\smallskip}\hline\noalign{\smallskip}
\end{tabular}
\end{table}
According to the proposed IFS approach, the results for \cub models in Table \ref{tab:mf} have to be read together with Table \ref{tab:nmf}, giving the corresponding non-membership functions \eqref{nmf}, and  with Table \ref{tab:uf}, giving the corresponding uncertainty functions \eqref{UncFunction}. 
\begin{table}[h!]
\caption{Non-membership function \eqref{nmf}}
\label{tab:nmf}       
\centering
\begin{tabular}[pos=c]{cccccc}
\hline\noalign{\smallskip}
Item &  $R = 1 $ & $ R = 2$ & $R = 3$ & $R = 4$ & $R \geq 5$ \\
\noalign{\smallskip}\hline\noalign{\smallskip}
\texttt{informat}  &  1 &  0.8230 &0.5569 &0.0295&   0 \\
\texttt{willing}   & 1 &0.8772 &0.4787 & 0.0205   & 0  \\ 
\texttt{officeho}  & 1 &0.7259 &0.5082 & 0.0457  & 0 \\
\texttt{compete}   &   1& 0.8305 &0.5085 & 0.0283  &  0 \\
\texttt{global}    &  1 & 0.8872 & 0.5057&  0.0188 &  0 \\
\noalign{\smallskip}\hline\noalign{\smallskip}
\end{tabular}
\end{table}

We notice that, the higher the value of $\hat{\pi}$ is, the more the non-membership degrees increase moving from the indifference point of the scale to its minimum. Instead, for  \texttt{officeho} (and in a lower measure for \texttt{informat}), the Fuzzy uncertainty function spreads more among all the categories and it is less concentrated around the indifference point.
\begin{table}[h!]
\caption{Fuzzy uncertainty function \eqref{uncertainty}}
\label{tab:uf}       
\centering
\begin{tabular}[pos=c]{cccccccc}
\hline\noalign{\smallskip}
Item &  $R=1$ & $R=2$ & $R=3$ & $R=4$ & $R=5$ & $R=6$ & $R=7$ \\
\noalign{\smallskip}\hline\noalign{\smallskip}
\texttt{informat} & 0 & 0.1770 & 0.4431 & 0.9410 & 0.6081 & 0.1770 & 0\\
\texttt{willing}  & 0 & 0.1228 & 0.5213 & 0.9591 & 0.6774 & 0.1228 & 0\\
\texttt{officeho} & 0 & 0.2741 & 0.4918 & 0.9086 & 0.6031 & 0.2741 & 0\\
\texttt{compete}  & 0 & 0.1695 & 0.4915 & 0.9435 & 0.6453 & 0.1695 & 0\\
\texttt{global}   & 0 & 0.1128 & 0.4943 & 0.9624 & 0.6523 & 0.1128 & 0\\
\noalign{\smallskip}\hline\noalign{\smallskip}
\end{tabular}
\end{table}

For the aggregation scores, Table \ref{tab:weights} compare the weights \eqref{pesiZani} both for the membership values \eqref{Zani} and for the uncertainty functions \eqref{uncertainty}. It turns out that the first weights do not suitably penalize the items \texttt{officeho} and \texttt{informat}, which should be given the least amount of importance since they register the highest uncertainty among the items (with reference to Table \ref{tab:param}, $1-\hat{\pi}=0.3198$ for \texttt{officeho} and $1-\hat{\pi}=0.2064$ for \texttt{informat}). Instead, for the weights based on the Fuzzy uncertainty function \eqref{uncertainty}, the lowest value is attained for item \texttt{officeho} ($w_3 = 0.1604$), and the weights values increase as the uncertainty $1-\hat{\pi}$'s decrease. This procedure takes into account also the level of feeling: indeed,  although item \texttt{willingn} is affected by a slightly higher uncertainty ($1-\hat{\pi}= 0.1433$) than \texttt{global} ($1-\hat{\pi} = 0.1316$), it is assigned a higher weight ($w_2 =0.2493$ for \texttt{willingn} against $w_5=0.2068$ for \texttt{global}) since to \texttt{willingn} it corresponds a larger feeling ($1-\hat{\xi}=0.8833$) than for \texttt{global} ($1-\hat{\xi}=0.8286$).
\begin{table}[h!]
\caption{The first two rows display the weights \eqref{pesiZani} computed with the Membership Function \eqref{Zani} and the Uncertainty Function
\eqref{uncertainty}. The last row reports the \cub uncertainty levels.}   
\label{tab:weights}       
\centering
\begin{tabular}[pos=c]{cccccc}
\hline\noalign{\smallskip}
  &  \texttt{informat} & \texttt{willingn} & \texttt{officeho} & \texttt{compete} & \texttt{global} \\
\noalign{\smallskip}\hline\noalign{\smallskip}
For Zani \textit{et al.} M.F. \eqref{Zani} &  0.2075 & 0.1713 & 0.2314 & 0.2006 & 0.1892 \\ 
For Fuzzy U.F. \eqref{uncertainty} &  0.1880 & 0.2493 &0.1604  & 0.1955 & 0.2068 \\ 
$1-\hat{\pi}$                      &  0.2064 & 0.1433 & 0.3198 & 0.1978 & 0.1316 \\ 
\noalign{\smallskip}\hline\noalign{\smallskip}
\end{tabular}
\end{table}

Finally, for both the weights summarized in Table \ref{tab:weights}, the final scores obtained via the IWAM aggregators \eqref{score} and \eqref{uncertaintyScore} are reported in the subsequent Table \ref{tab:finalscores}.
\begin{table}[h!]
\caption{Membership, non-membership and uncertainty scores}
\label{tab:finalscores}       
\centering
\begin{tabular}[pos=c]{cccc}
\hline\noalign{\smallskip}
Weights system \eqref{pesiZani} &  $\bar{\mu}$ & $\bar{\nu}$ & $\bar{u}= 1 - \bar{\mu} - \bar{\nu}$ \\
\noalign{\smallskip}\hline\noalign{\smallskip}
For Zani M.F. \eqref{Zani} &  0.5902 & 0.000 & 0.4098 \\
For Fuzzy U.F. \eqref{FuzzyPropUF} & 0.6669 & 0.066 & 0.2671 \\
\noalign{\smallskip}\hline\noalign{\smallskip}
\end{tabular}
\end{table}

\subsection{Comments and conclusions}
In conclusion, the fuzzification proposed in \eqref{CUBZani} behaves more efficiently compared with Zani \textit{et al.} approach since 
the aggregate uncertainty score $\bar{u} = 0.2671$ is reduced. This circumstance depends both on the different weighting of the indifference point by means of \cub model heterogeneity and on the membership and not-membership functions by means of the uncertainty of the distribution. A way to further reduce the final uncertainty score is to consider the \textit{shelter effect} \cite{Ian12a}, namely the occurrence of an inflated category, whose significance should be previously tested. We expect that the inclusion of a significant shelter effect in a positive category (to be tested for each item) adds more details on the feeling of the respondent when faces a questionnaire.
So the Fuzzy uncertainty score should reduce as  both the membership and non-membership scores increase. An in-depth analysis of this further step is left for future works. 


\begin{thebibliography}{99.}
%
\bibitem{Ferrari}
Andreis, F. and Ferrari, P.A. (2013) On a copula model with CUB margins. QdS Journal of Methodological and Applied Statistics, \textbf{15}, 33--51.
%
\bibitem{Cerioli}
Cerioli, A. and Zani, S. (1990) A fuzzy approach to the measurement of poverty, In: C. Dagum, M. Zenga (eds.), Income and Wealth distribution, Inequality and Poverty, Springer, Berlin, 272--284. 
%
\bibitem{Corduas}
Corduas, M. (2015) Analyzing bivariate ordinal data with CUB margins. Stat Modelling, \textbf{15}, 441--432.
%
\bibitem{Del00a}
D'Elia, A. (2000) A shifted Binomial model for rankings, Proceedings of the 15th International Workshop on Statistical Modelling. New Trends in Statistical Modelling (IWSM) - Bilbao, Spain,  412--416. 
%
\bibitem{DelPic2005}
D'Elia, A. and Piccolo, D. (2005) A mixture model for preference data analysis. Computational Statistics \& Data Analysis, \textbf{49}, 917--934.
%
\bibitem{VarCUB}
Gottard, A., Iannario, M. and Piccolo, D. (2016) Varying uncertainty in CUB models. Adv. Data Anal. Classif., 1--20, DOI 10.1007/s11634-016-0235-0 
%
\bibitem{Iannario} 
Iannario, M. (2012) Preliminary estimators for a mixture model of ordinal data. Adv. Data Anal. Classif., \textbf{6}, 163--184 
%
\bibitem{Ian12a}
Iannario, M. (2012) Modelling \emph{shelter} choices in a class of mixture models for ordinal responses. Stat Methods Appt., \textbf{21}, 1--22.  
%
\bibitem{Ianp12}
Iannario, M. and Piccolo, D. (2012)  CUB models: Statistical methods and empirical evidence. In: Kenett R. S. and Salini S. (eds.), Modern Analysis of Customer Surveys: with applications using R. Chichester: J. Wiley \& Sons, 231--258. 
%
\bibitem{CUBpackage}
Iannario, M., Piccolo, D. and Simone, R. (2015) CUB: A Class of Mixture Models for Ordinal Data (\textbf{R} package version 0.1), \url{http://CRAN.R-project.org/package=CUB}. 
%
\bibitem{Lalla} 
Lalla, M., Facchinetti, G. and Mastroleo, G. (2004) Ordinal Scales and Fuzzy Set Systems to Measure Agreement: An Application to the Evaluation of Teaching Activity, Qual. Quant., \textbf{38}, 577--601. 
%
\bibitem{Marasini} 
Marasini, D., Quatto, P. and Ripamonti, E. (2015) Intuitionistic fuzzy sets in questionnaire analysis. Qual. Quant.,  doi:10.1007/s11135-015-0175-3. 
%
\bibitem{Zadeh}
Zadeh, L.A. (1965) Fuzzy sets, Information and Control, \textbf{8}, 338--353.
%
\bibitem{Zani} 
Zani, S., Milioli, M.A. and Morlini, I. (2013) Fuzzy composite indicators: An applications for measuring customer satisfaction.
In: Torelli, N., Pesarin, F., Bar-Hen A. (eds.) Advances in Theoretical and Applied Statistics, 243-253. Springer, Heidelberg 
%
\end{thebibliography}
\end{document}